\documentclass[useASM]{mn2e}
\input{psfig.sty}
\usepackage{graphicx}

\newcommand{\be}{\begin{equation}}
\newcommand{\ee}{\end{equation}}
\newcommand{\ba}{\begin{eqnarray}}
\newcommand{\ea}{\end{eqnarray}}
\newcommand{\siml}{\lower4pt \hbox{$\buildrel < \over \sim$}}
\newcommand{\simg}{\lower4pt \hbox{$\buildrel > \over \sim$}}

\begin{document}

\title{On the Origins of Part Time Radio Pulsars}

\author[Zhang, Gil \& Dyks]
       {Bing Zhang$^{1}$, Janusz Gil$^{2}$ and Jaroslaw Dyks$^{3}$
 \\
$^1$ Department of Physics, University of Nevada, 
Las Vegas, NV 89154 \\ 
$^2$ Institute of Astronomy, Zielona G\'ora University,
Lubuska 2, 65-265 Zielona G\'ora, Poland \\
$^3$ Nicolaus Copernicus Center, Polish Academy of Sciences, Torun,
Poland }
\maketitle

\label{firstpage}

\begin{abstract}
Growing evidence suggests that some radio pulsars only act
sporadically. These ``part-time'' pulsars include long-term nulls,
quasi-periodic radio flares in PSR B1931$+$24, as well as the
so-called Rotating RAdio Transients (RRATs). 
Based on the assumption that these objects are isolated neutron stars
similar to conventional radio pulsars, we discuss two possible
interpretations to the phenomenon. The first interpretation suggests
that these objects are pulsars slightly below the radio emission
``death line'', which become active occasionally only when the
conditions for pair production and coherent emission are
satisfied. The second interpretation invokes a radio 
emission direction reversal in conventional pulsars, as has been introduced
to interpret the peculiar mode changing phenomenon in PSR B1822-09. In
this picture, our line of sight misses the main radio emission beam of
the pulsar but happens to sweep the emission beam when the radio
emission direction is reversed. These part-time pulsars are therefore
the other half of ``nulling'' pulsars. We suggest that X-ray
observations may provide clues to differentiate between these two
possibilities. 
\end{abstract}
\begin{keywords}
stars: pulsars - X-ray: observation - 
radiation mechanism: coherent - radio
\end{keywords}

\section{Introduction}
Recently, many interesting, peculiar radio-pulsar-like objects were
discovered.  It has been long known that old radio pulsars tend to
``null'' occasionally, i.e. the radio emission ceases occasionally for
one or more consecutive periods during the otherwise uninterrupted 
regular emission episodes (e.g. Ritchings
1976; Rankin 1986; Biggs 1992). Lewandowski et al. (2004) reported two
pulsars with very long period of nulling. While PSR J1649$+$2533
exhibits pulse nulling for approximately 30\% of the time, another
``bursting pulsar'', PSR J1752$+$2359, was found to have 70\%-80\% of
the time in the ``quasi-null'' state. The active and the null phases
of these pulsars usually last for 100s of periods. 
McLaughlin et al. (2006) reported the discovery of a new class of 
radio transients from the Parkes Multi-beam Pulsar Survey.  
The current sample includes 11
objects characterized by single, dispersed bursts of radio emission
with durations ranging from 2 to 30 milliseconds. Long-term monitoring
of these objects led to identifications of their spin periods ($P$),
ranging from 0.4 to 7 seconds. These fall into the range of the
$P$-distribution of the conventional radio pulsars, although on the 
long end. The period derivatives ($\dot P$) of three RRATs were 
measured, which are also typical for conventional 
pulsars. These objects on average have slightly higher brightness
temperatures than the conventional pulsar population, but the small
discrepancy is easily accounted for by an observational selection
effect (M. McLaughlin, 2006, personal communication). McLaughlin et
al. (2006) concluded that these objects represent a previously unknown
population of rotation-powered neutron stars, which they call Rotating
RAdio Transients (RRATs). A serendipitous {\em Chandra} observation has 
detected one of the RRATs (RRAT J1819-1458, Reynolds et al. 2006) in
X-rays, whose spectrum is dominated by a soft thermal component.
Lately, Kramer et al. (2006) reported that 
a previously known pulsar PSR B1931$+$24, when monitored long enough,
shows a very long-term, quasi-periodic flaring behavior. The
``on''-state lasts for 5-10 days, while the ``off'' state lasts for
25-35 days. More interestingly, the spindown rate is different during
the on and the off states, respectively. A further search for similar 
objects from the Parkes Multi-Beam Survey data revealed at least 
four more objects that share similar properties to PSR B1931$+$24 
(Lyne 2006). 

Understanding the physical origin of these ``part-time'' radio pulsars
as well as their relationship to the conventional radio pulsars is 
of great interest to understand the mechanisms of pulsar radio
emission. Here we discuss two possible interpretations to part-time
pulsars by assuming that they are isolated neutron stars similar to
conventional radio pulsars. We also discuss how X-ray data may be 
used to distinguish between these two possibilities.

\section{Model I: Re-activated dead pulsars}

It is believed that electron-positron pair production plays an
essential role in pulsar coherent radio 
emission. The condition for the failure of pair production usually
defines the pulsar radio emission ``death line'' in the $P-\dot P$
diagram of pulsars\footnote{The real condition for pulsar coherent
radio emission may be more stringent than this, since it invokes
additional criteria on how coherence is generated and maintained. So
the pair production condition is only a necessary but not a sufficient
condition for pulsar radio emission.}. Although the death
line issue has been re-investigated many 
times over the years (e.g. Ruderman \& Sutherland 1975; Chen \& Ruderman
1993; Arons 2000; Zhang et al. 2000; Hibschman \& Arons 2001; Gil \&
Mitra 2001; Harding \& Muslimov 2002; Harding et al. 2002), it is very
difficult and essentially impossible to define an exact line in the
$P-\dot P$ diagram. This is due to many uncertainties inherited to
the death line problem (Zhang 2003), including the criterion to define
pulsar death (whether the pair multiplicity is zero or only a small
factor with respect to the primary particle number density, say, less
than unity), the boundary condition of the inner ``gap'' (a vacuum
gap, a space-charge-limited flow - SCLF, or a partially screened
gap); the gamma-ray emission mechanism (curvature radiation or inverse
Compton scattering), the equation of state of the neutron star (or
even strange quark star), and especially the unknown near-surface
magnetic field configuration. What is relevant would be then a
so-called ``death valley'' in the $P-\dot P$ diagram (named by Chen \&
Ruderman 1993). Pulsars in the valley could either be ``alive'' or
``dead'' according to the properties of the individual pulsars.

Recently, the enigmatic radio bursting source in the
direction of Galactic Center GCRT J1745$-$3009 (Hyman et al. 2005)
was intepreted by Zhang \& Gil (2005) as a transient white dwarf
pulsar with a rotation period of $77.13$ minutes. Assuming a dipolar 
surface magnetic field $\sim 10^9$ G, GCRT
J1745-3009 is slightly below the white dwarf pulsar ``death line''. 
Zhang \& Gil (2005) then argued that if stronger multipole magnetic
fields emerge to the polar cap region of the white dwarf, the pair
production condition could be satisfied and the white dwarf could then
behave like a neutron star radio pulsar. This corresponds to the
observed 5 consecutive outbursts at 0.33 GHz with a $77.13$-minute 
period and a $10$-minute duration for each outburst recorded during 
2002 Sep. 30th to Oct. The putative strong magnetic fields
may not last long. This accounts for the cessation of the radio
emission from the source. The reactivation of the source in 2003 
(Hyman et al. 2006) corresponds to another episode of pair
production in the white dwarf pulsar.

A natural inference from the above suggestion is that one would also
expect similar situations for neutron star pulsars. Radio pulsars may
not die abruptly at the end of their lives. Rather, they may
experience a period of life during which the pair production processes
could be maintained only sporadically. In other words, many pulsars
not deep below the death line could jump out from the graveyard
occasionally, if the pair production condition could be temporarily 
satisfied occasionally. One possibility is that strong sunspot-like 
magnetic fields emerge into their polar cap regions. Within such a
scenario, part-time pulsars are simply these
not-quite-dead (zombie) pulsars. The measured $P$ for RRATs is typically 
long (5 out of 10 have $P>4$ s, McLaughlin et al. 2006). This makes
them more likely to locate in the death valley. For example, RRAT
J1317$-$5759 ($P = 2.64$ s, $\dot P = 12.6 \times 10^{-15}$,
MaLaughlin et al. 2006) is below the curvature radiation death line
for a pure star-centered dipolar field, i.e. $\log \dot P = (11/4)
\log P -14.62$ for a vacuum gap and $\log \dot P = (5/2) \log P
-14.56$ for a SCLF (Zhang et al. 2000).
If the near-surface field configuration is nearly dipolar
and if curvature radiation is the dominant mechanism to
produce enough pairs to power radio emission, RRAT J1317$-$5759 should
be radio quiet most of the time since it is below the death line. If
stronger sunspot-like fields merge into the polar cap region, the pair
production condition could be temporarily satisfied, so that a
part-time pulsar could be temporarily re-activated. 

Not all part-time pulsars are below the star-centered dipolar death
line. For example, RRATs J1819$-$1458 ($P=4.26$ s, $\dot
P = 576 \times 10^{-15}$) and J1913$+$1333 ($P=0.92$ s, $\dot P = 7.87
\times 10^{-15}$) (MaLaughlin et al. 2006) are both somewhat above
the star-centered dipolar death line. The spin parameters of 
PSR J1752$+$2359 ($P=0.41$ s, $\dot P=0.64 \times 10^{15}$,
Lewandowski et al. 2004) and PSR B1931$+$24 ($P=0.81$ s, $\dot P=8.11
\times 10^{15}$, Kramer et al. 2006) also place them slightly above
these death lines.  In order to interpret them as re-activated 
pulsars, one needs to argue that in general the near-surface
magnetic field strengths of these objects are overestimated. This
could be simply due to the inaccuracy of the estimates introduced
by a crude dipole spindown model. Alternatively, this could be caused
by an off-center (but still axisymmetric) dipolar magnetic field
configuration of the neutron star (e.g. Arons 2000). In such 
a picture, the ``near-end'' polar cap has a stronger magnetic field
than the case of a star-centered dipole, while the ``far-end'' polar
cap has a weaker field. While some active pulsars deep below the
star-centered dipolar death line may be those cases we see the
near-end polar caps, part-time pulsars would be those we see the 
far-end polar caps. Within such a scenario, the part time pulsars
are not systematically older than other pulsars in the death
valley; their peculiar behavior is caused by their unfavorable 
viewing geometry, i.e., one sees the far-end off-center dipole where
the local magnetic fields are systematically weaker than their 
brethren. For example, the thermal X-ray emission RRAT J1819$-$1458
is bright (Reynolds et al. 2006), which is consistent with being a 
young pulsar. The pair production condition could be however not 
satisfied most of the time, if our line of
sight sweeps a far-end dipole where the near-surface magnetic field
is too weak.

Another source of gamma-rays to trigger pair production is inverse
Compton scattering (IC, e.g. Zhang \& Qiao 1996). For star-centered 
dipoles, some RRATs are above the death line of resonant-IC-controlled
gaps (both vacuum gaps and SCLFs, Zhang et al. 2000), and all part
time pulsars are above the death line of non-resonant-IC-controlled 
SCLFs (Harding et al. 2002). However, detailed modeling suggests that 
pair multiplicity could be much lower in IC-controlled SCLFs than in 
CR-controlled SCLFs for a star-centered dipole (Harding \& Muslimov
2002; Harding et al. 2002). It is unclear whether the resulting low
pair multiplicity could be sufficient to trigger strong coherent radio
emission in these pulsars. Since there is no abrupt change of radio
emission properties for pulsars across the star-centered CR death
lines, one could speculate that there might not be two distinct types
of pulsars controlled by CR and IC, respectively. The reactivation
model relies on the conjecture that pulsar pair cascades are induced
by curvature radiation only.

The existence of multipole magnetic fields near pulsar polar caps has
been required by the earliest pulsar models (e.g. Ruderman \&
Sutherland 1975). Theoretically strong spot-like magnetic fields
could be generated from the sub-surface toroidal magnetic field 
component through Hall-drift induced instability (Gepper et al. 2003) 
or through small scale turbulent dynamo actions 
(Urpin \& Gil 2003). During the spin-down of a pulsar, the magnetic
flux is expected to move as a consequence of the interaction between
neutron and proton superfluid vortices (Ruderman 1991; Ruderman et al.
1998; Jones 2006). The dynamical evolution of the field configuration 
may be slow. The pair production condition, on the other hand, has a 
threshold (Ruderman \& Sutherland 1975),
i.e. $(E_{ph}/2m_e c^2)(B_\perp/B_q) \sim 0.1$ (Ruderman \&
Sutherland 1975), where
$E_{ph}$ is the photon energy, $B_\perp$ is the perpendicular magnetic
field encountered by the photon, and $B_q=4.414\times 10^{13}$G is
the critical magnetic field. 
A pair production cascade is abruptly developed as soon as the 
threshold condition is met. The dynamical time scale of the inner gap 
($\sim h_{\rm gap}/c \sim 10^{-6} - 10^{-4}$ s, where $h_{\rm gap}$ 
is the height of the gap) is
much shorter than the rotation period $P$. So the time scale to turn
on the pulsar radio emission is typically much shorter than $P$. 
The time scale during which the pair condition is satisfied is hard
to derive from the first principles. 
From the data, it seems that this time scale varies in a wide range.
Most RRATs have one single pulse in each burst, with one having 
multiple periods within a single burst (M. McLaughlin 2006, personal 
communication). Long-term nulling pulsars (e.g. PSR J1752$+$2359, 
Lewandowski et al. 2004) sustain the radio emission for 100s of 
periods, while PSR B1931$+$24's on-state lasts for 5-10 days 
(Kramer et al. 2006). 
The strength of the evolving multipole fields is not constrained, 
but should be around $10^{12}$ G or higher in order to make the model
work. The dynamical evolution of the fields 
should occur throughout the neutron star's life time. However, in
young pulsars with small $P$, the contribution of these evolving 
fields may not be significant since the stable field lines have 
a large enough curvature to facilitate pair production. Only in slow
pulsars whose stable $B_\perp$ component is small enough do the 
evolving components dominate the pair production process. This is 
consistent with the fact that part-time pulsars tend to have
long periods.

A direct consequence of the reactivation model is that the global
pulsar current is turned on only during the reactivated phase. A clear
change of spindown torque in PSR B1931$+$24 during the ``on'' and
``off'' phases (Kramer et al. 2006) lends strong support to this
scenario.  

\section{Model II: nulling pulsars viewed at the opposite direction}

The measured spin parameters of the three RRATs (J1317$-$5759,
J1819$-$1458 and J1913$+$1333, McLaughlin et al. 2006) and other
part-time pulsars (PSR J1752$+$2359, Lewandowski 
et al. 2004; and PSR B1931$+$32 Kramer et al. 2006) do not differ 
significantly from
those of conventional pulsars. This raises the possibility 
that these objects are intrinsically similar to conventional pulsars
but appear differently because of certain geometrical reasons. One
possible picture is the emission direction reversal mechanism
recently proposed by Dyks et al. (2005a) to interpret the peculiar
mode-changing behavior of PSR B1822$-$09 (Gil et al. 1994). 

Figure 4b from Gil et al. (1994) displays an interesting mode
switching phenomenon for PSR B1822$-$09. There are three emission
components located at phases $17^{\rm o}$, $33^{\rm o}$, and $215^{\rm
o}$, respectively. The first two peaks are termed as the main pulse,
and the third one is called the interpulse. While the second peak of
the main pulse appears all the time, there is an apparent switching
on/off anti-correlation between the first peak of the main pulse and
the interpulse, i.e. the first peak is on whenever the interpulse
is off, and vice versa. Such a phenomenon has been difficult to
interpret within the traditional pulsar models. Dyks et al. (2005a)
proposed that pulsar 
radio emission may occasionally reverse direction. According
to this hypothesis, PSR B1822$-$09 is a special case in which our line
of sight happens to sweep the emission beams of both the traditional
outward emission and the inward emission during the reversal phase.

A natural inference from the reversal hypothesis is that in most
geometric configurations, the line of sight can only sweep one emission
beam, either the outward main beam or the inward one. Dyks et
al. (2005a,b) proposed that pulsar ``nulling'' is caused by reversal,
and that the conventional nulling pulsars are those pulsars whose outward
main pulse is swept by the line of sight. Within this picture, there
should be also cases when only the reversed inward component is
seen. These objects would be identified as 
part-time pulsars. As a result, part-time pulsars are the
``opposite'' population of nulling pulsars.

The origin of the emission direction is unknown. One possibility 
may be the large-amplitude oscillation of the current far above 
the surface (Levinson et al. 2005). This is preferably achieved for a
charge-starving intial condition (Levinson et al. 2005), which tends
to happen below the curvature emission death line if the near
surface magnetic field is nearly dipolar (Harding \& Muslimov 2002).
This would be consistent with that part-time pulsars tend 
to concentrate in the
death valley. According to this picture, pulsar death is not only
associated with the inability of producing pairs, but the processes
such as current oscillations may be also relevant.  
In the case of PSR B1822$-$09 (Gil et al. 1994), a sequence of 
pulses appear at the interpulse phase when the proposed reversal 
occurs. This would be consistent with the bursting pulsar
PSR J1752$+$2359 (Lewandowski et al. 2004). RRATs usually show 
one pulse during each burst. On the other hand, many nulling pulsars 
only miss one pulse during each null. 

Since the reversal model interprets nulls and bursts of radio emission
via a geometric effect, no significant spindown torque change is
expected during the ``on'' and ``off'' states. This model is therefore
ruled out at least for PSR B1931$+$32 (Kramer et al. 2006).

\section{X-rays as possible differentiator}

The two models suggested in this paper are plausible ways of 
interpreting the available radio emission data of
part-time pulsars. It would be interesting to find some criteria
to differentiate between the two mechanisms. We suggest that
X-ray data may provide important clues. 

In general the X-ray emission of a neutron star consists of three
possible components, although not all the three components are
detectable in every pulsar: (1) a pulsed, non-thermal component
originating from the magnetosphere, (2) a thermal component from the
bulk of the star due to neutron star cooling, and (3) a 
hot thermal component from a small area on the neutron star surface,
possibly due to enhanced heating at the polar cap. Although
the cooling component is determined by age only, and hence, cannot
be used to differentiate between the two scenarios, the other two 
emission components are potentially useful to put constraints on the
above-mentioned two scenarios.

In the first (reactivation) model, no significant X-ray
emission from these two components are expected in the quiescent
state, during which the magnetosphere is charge-starved. 
There could still be a
primary particle outflow, but the radiated energy is likely in the
gamma-ray band (e.g. Muslimov \& Harding 2004). One therefore does not
expect a strong non-thermal X-ray component of magnetospheric origin 
during the quiescent state. For the same
reason, one does not expect a strong active returning particle flow,
and hence, a significantly heated polar cap. Potentially
during the reactivation phase, the resumed magnetospheric activities
would lead to both non-thermal X-ray emission and polar-cap
heating. A coordinated simultaneous observation in both X-ray and
radio bands is therefore desirable to test this scenario. 
Because the ``duty cycles'' of RRATs' activity are
very short (McLaughlin et al. 2006), the enhancement of X-ray
emission during the active phase of RRATs may be too small to be
observed with the current X-ray telescopes. The bursting pulsar
PSR J1752$+$2359 and the flaring pulsar PSR B1931$+$32, on the other 
hand, could be ideal sources to perform such a test.

The second (reversal) model interprets part-time pulsars as nulling
radio pulsars viewed at the opposite direction so that the reversed 
emission beam is detected. Nulling pulsars are usually middle-aged  
to old pulsars (e.g. Table 4 of Rankin 1986). A growing sample of
pulsars in this age-group have been observed by X-ray observatories
({\em Chandra} and {\em XMM-Newton}, e.g. Becker et al. 2004, 2005;
Zavlin \& Pavlov 2004; Zhang et al. 2005; Tepedelenliolu \& \"Ogelman
2005; De Luca et al. 2005; Kargaltsev et al. 2006). These observations
indicate that the emission of these pulsars likely has multiple
emission components, including a non-thermal magnetospheric component
and sometimes a hot-spot thermal emission component. The latter has 
been claimed to be discovered in several nulling pulsars in Table 4 of
Rankin (1986), including PSR 0628-28 (Tepedelenliolu \& \"Ogelman
2005), PSR 0656+14 (De Luca et al. 2005), and PSR 1133+16 (Kargaltsev
et al. 2006). This component is likely produced by the returning
particle flow that precipitates and heats the polar cap region 
(e.g. Zhang et al. 2005; Gil et al. 2006 for more discussion).

Generally part-time pulsars would display the similar X-ray emission
properties to the nulling pulsars within the reversal model, since
the underlying pulsars are supposed to be active. The observed
X-ray spectrum should then generally include a magnetospheric
component and/or a hot spot component. There may be exceptions from
this rule because the significance of these spectral components
depends on the viewing geometry. For example, the projected area and
luminosity of the hot spot would be diminished if the reversed viewing
direction is far off the magnetic poles. The direction of non-thermal
X-rays in the magnetosphere (either outwards, Zhang \& Harding 2000; or
inwards, Cheng et al. 1998; Wang et al. 1998) is unknown, so that the
detectability and significance of this component in the reversed
geometry is uncertain. Nonetheless, detecting either the hot spot
component or the magnetospheric component would be a strong support to
the reversal scenario. 

To summarize, detailed X-ray observations should shed light onto
the nature of part-time pulsars. If a strong
non-thermal emission component and/or a distinct ``hot spot'' thermal
component are detected during the ``off'' mode of radio emission, the
reactivation model (model 1) is essentially rejected, and the reversal
model (model 2) is generally supported. Non-detections, on the other
hand, would be consistent with the reactivation model and disfavor the
reversal model, although more detailed modeling is needed to tell 
whether the model is completely ruled out.

RRAT J1819-1458 was serendipitously detected with {\em Chandra} by
Reynolds et al. (2006, see also Gaensler et al. 2006). The spectrum 
of the X-ray counterpart is similar to that of a radio
pulsar of comparable age, which is dominated by 
a soft thermal component due to neutron star cooling. No strong
non-thermal magnetospheric emission nor a hot-spot thermal emission 
were identified. The data are therefore consistent with the
reactivation pulsar model, and the reversal model
is disfavored with the present quality of the X-ray spectrum.

\section{Conclusions and discussion}

We have suggested two possible interpretations to the recently 
identified part-time pulsars. One is that these objects are
re-activated dead pulsars slightly below the conventional radio
emission (pair production) death line. The other is that they are
simply the other half of nulling pulsars for which our line of sight
misses the outward-directed main radio emission beam but happens to
sweep the reversed inward-directed emission beam. 

Since the predicted X-ray emission properties differ significantly
from each other in the two scenarios, we suggest an X-ray diagnostic
to differentiate between the two possible interpretations. 
In particular, a coordinated X-ray and radio observational campaign 
would be essential to unravel the nature of these part-time pulsars. 
If a strong non-thermal magnetospheric emission component 
and/or a hot-spot thermal emission component are identified from their
X-ray spectra, these objects are then very likely similar to
conventional pulsars. The emission direction reversal and a preferred
viewing geometry are likely the agents to make a part-time
pulsar. Such an identification would lend support to the inward
emission proposal for radio pulsars (Dyks et
al. 2005a,b). Alternatively, if after deep searches no strong
magnetosphere-related X-ray emission components (non-thermal or hot
spot thermal) are detected from any of these objects, part-time
pulsars are then very likely not-quite-dead pulsars before
disappearing in the graveyard. This would suggest that the
microscopic condition near the pulsar polar cap region is much more
complicated than usually imagined. This allows us to directly study
the dynamical evolution of the magnetic fields near the polar cap
region. In any case, either possibility would provide profound
implications for understanding the poorly known pulsar radio emission
mechanism. 

So far the available data seem to be consistent with the reactivation
model, while the reversal model is disfavored for at least PSR
B1931$+$24 and RRAT J1819$-$1458. 
A systematic search for X-ray counterparts of RRATs and other
part-time pulsars is desirable to draw a firmer conclusion.

Several other suggestions have been proposed recently to interpret
part-time pulsars.

1. Popov, Turrola \& Possenti (2006) suggest that RRATs may be related
to X-ray dim isolated neutron stars (XDINSs) based on comparisons of 
the birth rate and X-ray property of the two populations. The 
reactivation model discussed here would offer a mechanism to generate 
part-time pulses from these objects. If it turns out that all RRATs 
are X-ray dim, the reversal model would be disfavored since it 
predicts bright non-thermal and hot-spot thermal emission components. 
The discoveries of these components, on the other hand, would disfavor 
the suggestion that RRATs are related to XDINSs. 

2. Cordes \& Shannon (2006) and Li (2006) independently suggest that
part-time pulsars are modulated by sporadic accretion into the
neutron star magnetosphere that quenches the coherent radio emission.
An earlier suggestion along the same line has been made by Wright 
(1979). In these scenarios, there is another X-ray emission component 
due to accretion. However, a proper adjustment of parameters 
(e.g. Li 2006) would make the model satisfy the X-ray constraint 
from RRAT J1819-1458 (Raynolds et al. 2005). Based on X-ray data 
alone, it is not easy to differentiate these models from the ones 
we propose.

3. Weltevrede et al. (2006) argue that bright pulses detected in
pulsars such as PSR B0656$+$14, if the pulsar is far away enough, 
would mimic RRATs' emission. In view that the X-ray spectrum of 
RRAT J1819$-$1458
(dominated by a cooling thermal component, Reynolds et al. 2006) 
is not fully consistent with that of PSR B0656$+$14 (with the 
existence of another ``hot spot'' thermal component and possibly 
a non-thermal component besides the cooling thermal component, 
e.g. Marshall \& Schulz 2002; De Luca et al. 2005), this 
suggestion may not be conclusive. Even if
RRATs are B0656$+$14-like objects, some other part-time pulsars 
still call for other interpretations including the ones suggested 
in this paper.

We thank two referees for helpful comments. BZ thanks Maura 
McLaughlin for informative communication, Dick
Manchester, Alice Harding and Brien Gaensler for helpful discussion. 
This work is supported by NASA under grants NNG05GB67G (BZ) and by
grant 1 P03D 029 26 of the Polish State Committee for Scientific
Research (JG).

\end{document}